# Oblique Angle Deposition of $HfO_2$ Thin Films: Investigation of Elastic and Micro Structural Properties

By

R B Tokas[1], S Jena[1], P Sarkar[1], Shyam polaki[2], S Thakur[1], S Basu[3] and N K Sahoo[1]

[1]Atomic & Molecular Physics Division, Bhabha Atomic Research Centre, Trombay, Mumbai

[2]Indira Gandhi Centre for Atomic Research, Kalpakkam

[3]Solid State Physics Division, Bhabha Atomic Research Centre, Tromaby, Mumbai

**Abstract**

Oblique angle deposition of oxides is being very famous for fabricating inhomogeneous thin films with variation of refractive index along thickness in a functional form. Inhomogeneous layers play a key role in the development of rugate interference devices for photo-physical applications. Such obliquely deposited thin films show high porosity which is a critical issue related to their mechanical and environmental stability. Hence, it is important to investigate elastic properties of such film in addition to optical properties. Using atomic force acoustic microscopy, we report indentation modulus of $HfO_2$ thin films deposited at angles 80, 68, 57, 40 and 0 degree with normal to substrate plane on Si (100) substrate. Such films were measured to have indentation modulus of 42 GPa for extreme obliquely deposited film and indentation modulus increases with decrease in angle to become highest with a value of 221 GPa for normally deposited films. We also report microstructural properties and density of films measured by FESEM and grazing angle X-ray reflectometer respectively. Both indentation modulus and density depict a parabolic decreasing behavior with angle of deposition. Variation of density is again confirmed by FESEM cross-sectional morphology of such films.

**Introduction**

In recent years oblique angle deposition has been drawing attention due to applications in photonic crystals, optical interference devices, micro censors, microelectronics and rugate interference filters. Now a day's oblique angle deposition of thin films is being famous for fabricating rugate interference filters using single optical materials by varying refractive index along the film thickness [1-3]. This refractive index variation is achieved by changing the angle of deposition and is the result of varying porosity due to atomic shadowing and limited ad-atom diffusion [4-7] during growth. Oblique angle deposition generally works at an angle greater than 60 degree with normal to substrate. When this angle reaches around 80 degree, it is called glancing angle deposition (GLAD). Oblique angle deposition results in special morphological nanostructures & microstructure and by employing substrate rotation and varying deposition angle, pillar, helix, zigzag, erect columns etc. have been achieved successfully [8-11].Yumei Zhu et al., have fabricated multi-stop band interference rugate filter exploiting GLAD technique [12]. Stephan Fahr et al., have developed optical rugate filters for light trapping in solar cells [13]. Researchers have also developed GLAD antireflection coating [14], selective polarization transmission filter [15], narrow band pass rugate filter [16] and relative humidity sensors [17] by exploiting oblique angle deposition of optical materials.

Elastic properties are important and indispensable parameters in the assessment of environment & mechanical stability [18] of thin films and differ significantly from those of the bulk materials due to the interfaces, microstructure, and the underlying substrates. Like optical and structural properties, elastic properties such as the stress, co-efficient of thermal expansion, poison's ratio and indentation modulus of the film are also affected by the micro or nano-structure, the deposition process parameters and the deposition technique [19].Oblique angle deposited thin films are highly porous due to engineered microstructure and hence their mechanical stability is a subject of concern. Therefore it is of great importance to characterize and optimize elastic properties of porous thin films. Many techniques are being used for the determination of elastic properties of thin films including micro-and nano-indentation tests [20], laser induced surface acoustic wave (SAW) [21], Surface Brillouin light scattering measurements [22] and Atomic Force Acoustic Microscopy [23]. The nano indentation techniques have limitations due to slowness, limited lateral resolution and substrate effect [24], and also the technique is inherently destructive [25]. The laser induced SAW method needs rigorous analysis and also has limitations in terms of low sensitivity and limited spectrum band width [26]. Surface Brillouin light Scattering is very time consuming and sensitive to the environment noise [27]. Atomic Force Acoustic Microscopy (AFAM) technique, on the other hand, can give both quantitative values of Young's modulus of thin films like other technique as well as qualitative picture of Young's modulus in terms of acoustic image simultaneously with surface topography of thin film without being affected by the elastic properties of the substrate [28]. This makes AFAM, a unique and attractive technique for the investigation of elastic properties of thin films. AFAM technique is based on the evaluation of contact stiffness between tip and sample [29]. Contact resonance frequency for I, II and if possible III overtones are compared to free resonance frequency cantilever probe. From the shift of contact resonance frequencies from free vibrations, the contact stiffness (tip-sample) is computed by solving tip-sample vibration characteristic equation [29-32]. Same measurement is carried out on reference material with known indentation modulus. By employing Hertzian contact formulation and indentation modulus of reference material, indentation modulus of test sample is computed. In AFAM measurement technique, tip geometry plays an important role in determining the indentation modulus. For any AFM based technique used to quantify Indentation modulus of the sample, quantitative elastic properties and geometry of cantilever tip are required. Direct measurement of elastic properties of probe is a very difficult & erroneous and researchers assume bulk values for the probe. Estimation of indentation modulus with such assumptions and single reference material leads to uncertainty more than 20% [33-34]. Rabe et al. [35-36], proposed dual reference method to determine indentation modulus of unknown sample. This method does not rely on tip geometry & tip elastic properties and can offer results with uncertainty as low as 1%, depending on selection of references.

Hafnium oxide is a widely used optical material which possesses high refractive index and high laser induced damage threshold [37-40]. Its wide band gap (∼5.5-5.7eV) [37, 41] gives its transparency over a wide spectral range, extending from the ultraviolet to the mid-infrared [42].In present study, we have employed dual reference method to estimate indentation modulus of obliquely deposited $HfO_2$ thin films electron beam evaporation. We have also performed grazing incidence X-ray reflectivity (GIXR) measurements to determine the film density and RMS surface roughness. Cross-sectional morphology of such films has also been measured by field emission scanning electron microscopy (FESEM) to explore the microstructure and to find thickness. A correlation among indentation modulus, film density and microstructure of films deposited at different angles has been set.

## 2. Experimental Detail

### 2.1. Oblique angle deposition of $HfO_2$ thin film

In present wok, five $HfO_2$ thin films have been deposited on silicon (100) substrate at $200^0C$ by reactive electron beam evaporation technique in a VTD make 'Vera-902" EB evaporation coating system.Before deposition entire Si substrates were cleaned ultrasonically to achieve good quality films. Such films were deposited at angles (α) = 0, 40, 57, 68 and 80 degree and films have been designated as SAMP-5, SAMP-4, SAMP-3, SAMP-2 and SAMP-1 respectively. The angle of deposition (α) is defined as the angle between normal to the substrate plane and incident vapor flux. Different values of angle were set by tilting the substrate whereas incoming vapor flux was held fix. The base pressure prior to deposition was kept $1x10^{-5}$ mbar and. During deposition, high purity (99.9%) oxygen was supplied in to vacuum system through mass flow controller to maintain the stoichiometry of $HfO_2$ thin films and an optimized oxygen partial pressure was kept at $1x10^{-4}$ mbar. Rate of deposition and film thicknesses were monitored and controlled by Inficon make 'XTC2' quartz crystal micro-balance. Deposition rate was maintained to be 5Å/s.

*2.2. AFAM Characterization: measurement of contact resonance frequencies*

In contact resonance measurements, the sample under investigation was bonded to a piezoelectric transducer placed on AFM stage just below the sample. Piezo-transducer generates acoustic waves from 0.1-5.0 MH. Honey being an excellent ultrasonic transmitter has been used to couple the sample to ultrasonic transducer. Ultrasonic transducer was excited with a continuous sine wave generator from 0.1 to 5 MHz frequency and 0-1 V signal. Schematic of AFAM set-up utilized for our experiments is described in Fig. 1 (a). The cantilever probe was brought to sample contact in repulsive mode and $I^{st}$ & $II^{nd}$ contact resonance frequencies were measured at many different points on $HfO_2$ thin film and mean contact resonance frequency was used for further calculations. Contact resonance frequencies measurement was also performed on reference samples viz., BK7 glass, Silicon (100) and Sapphire, whose indentation modulii have been taken as reference. Two references together were used to eliminate the requirement of geometry and indentation modulus of measuring cantilever probe for the computation of indentation modulus of thin films. Contact resonance measurement was carried out using NT-MDT, Russia make P47H system and diamond like carbon (DLC) coated tip on Si cantilever has been used for entire experiment. DLC coated tip was chosen because of its long durability, high hardness and high resistance to the changes in geometry because of measurements. In AFAM, a laser beam deflection feedback mechanism is employed which controls the force between the tip and sample.

*2.3. Grazing angle X-ray reflectivity (GIXR) and cross-sectional morphology measurements*

Density of the films has been estimated from GIXR measurements carried out in X-ray reflectometer. The measurements have been carried out with a Cu $K_\alpha$ (1.54Å) source with grazing angle of incidence in the range of 0-0.5° with angular resolution of 0.01°. The detailed theory of GIXR measurements of thin films is discussed in [43].

X-ray suffers total external reflection at extreme grazing angle of incidence from any surface. However, as the grazing angle of incidence value ($\theta$) exceeds the critical angle ($\theta_c$), X-ray starts penetrating inside the layer and reflectivity falls rapidly. The critical angle is approximated by:

$$\theta_c = \sqrt{(2\delta)} \qquad (1(a))$$

Here $\delta$ is a function of electron density, wavelength etc. The reflectivity of X-ray from a thin film i.e., of a plane boundary between two media can be obtained using the well-known Fresnel's boundary conditions of continuity of the tangential components of the electric field vector and its derivative at the sharp interface [43]. However, the Fresnel's reflectivity gets modified for a rough surface by a 'Debye-Waller -like' factor as follows:

$$R_o = R_p \, exp \, (-q^2 \sigma^2/2) \qquad (1(b))$$

where, $q$ is the momentum transfer factor ($4\pi \, Sin\theta/\lambda$), Ro is the reflectivity of the rough surface and Rp is the reflectivity of an otherwise identical smooth surface and $\sigma$ is the RMS roughness of the surface. Thus by fitting the X-ray reflectivity spectrum of the surface of a sample near its critical angle, accurate estimation regarding the density $\rho$ and RMS surface roughness $\sigma$ can be made quite accurately. Cross-sectional morphology of the samples was recorded by field emission scanning electron microscopy (FESEM), Zeiss Supra 55VP system.

## 3. Computation of Indentation Modulus from AFAM measurements

AFAM technique and detailed methodology of contact resonance frequency measurement have been discussed extensively in references [34-35]. In order to relate the measured contact resonance frequency quantitatively to the sample elastic properties, cantilever can be modeled as rectangular uniform beam. Contact stiffness $k^*$ is evaluated from contact resonance frequency using the characteristic equation for test sample-cantilever coupled vibrations. The derivation of characteristic equation from equation of motion of flexural vibrations as well as torsional vibrations of cantilever tip has been discussed in references [30-33]. The forces like elastic, adhesion, friction always play a significant role when cantilever tip comes in contact with sample surface. Such forces are nonlinear function of distance between the tip and the sample surface. Analytical solution of such equation of motion is a rather cumbersome task. But by assuming forces linear for small vibration amplitude, tip-sample interaction can be approximated by vertical & lateral spring dashpot system [35]. The total length of cantilever can be defined $L = L_1 + L'$. $L_1$ is the actual position of tip position from cantilever base. It is well reported that [33, 44-45] normal contact stiffness $k^*$ depends much more on the sensor tip position $L_1$ than on any other parameters such as tip-surface angle differing from normal, lateral contact stiffness and lateral damping constant & air damping. In a general case, by neglecting tip mass and above mentioned parameters, tip-sample coupled system reduces to a very simple system shown in Fig. 1(b). In the simplest model, only vertical contact stiffness $k^*$ representing the interaction forces describes the vibrations of the surface coupled cantilever. The static normal load, significantly greater than the adhesion forces are needed to apply to the cantilever in order to neglect the adhesion forces effect on tip-sample interaction. In this way, the elastic forces become much dominating in tip-sample interaction and hence tip-sample normal contact stiffness. The characteristic equation for normal contact stiffness for simplified system is as following [35, 46]

$$k^* = \frac{\left(\frac{2}{3}\right) k_c (k_n L r)^3 (1 + \cos k_n L \cosh k_n L)}{\begin{bmatrix} \{(\sinh k_n L r \cos k_n L r - \cosh k_n L r \sin k_n L r)(1 + \cos k_n L(1-r) \cosh k_n L(1-r))\} \\ + \begin{Bmatrix} (\cosh k_n L(1-r) \sin k_n L(1-r) - \sinh k_n L(1-r) \cos k_n L(1-r)) \\ (1 - \cos k_n L r \cosh k_n L r) \end{Bmatrix} \end{bmatrix}} \quad (2)$$

Where $r = L_1/L$, $k_n$ is wave number of $n^{th}$ eigen mode and $k_c$ is the spring constant or stiffness of rectangular cantilever. To determine the value of $k^*$ from equation (2), values of $k_c$, $k_n L$ and $L_1$ are required. Value of $k_c$ can be determined from Sadar normal method [46] by using free cantilever resonance frequency spectra which are achieved by exciting cantilever by a piezo-frequency generator. The resonance frequencies $f_n$ of the cantilever being related to wave number $k_n$ are given by [35]

$$k_n L = C_B L \sqrt{f_n} \quad (3)$$

Where $k_n = \frac{2\pi}{\lambda_n}$ and $\lambda_n$ is acoustic wavelength. $C_B$ is the cantilever characteristic constant and depends on density, Modulus of elasticity (E) and geometry of cantilever. For free vibration of condition of cantilever (when tip is far from sample surface), tip-sample contact stiffness ($k^*$) is zero and for this condition equation (2) reduces to

$$1 + \cos k_n L \cosh k_n L = 0 \quad (4)$$

This is the characteristic equation for free flexural vibrations of cantilever. For different vibration mode (n), solutions of equation (3) are as following [31]

$$k_1 L = 1.8751, k_2 L = 4.69, k_3 L = 7.85 \ldots\ldots\ldots \quad (5)$$

By putting the values of $kL$ and resonance frequency $f$ in equation (3), the value of $C_B L$ can be obtained precisely without knowing the mechanical or elastic properties of cantilever. In order to determine the actual tip position and corresponding contact stiffness, we use the following criteria

$$k^*(f_1, L_1/L) = k^*(f_2, L_1/L) \tag{6}$$

Here $f_1$ & $f_2$ are $1^{st}$ and $2^{nd}$ contact resonance frequencies. The solution of above equation is determined numerically by taking range of $L_1/L$ from 0.85 to 0.98. For most of the cantilevers, the value $L_1/L$ lies in this range. From equation (6), the value of $L_1/L$ is evaluated and by substituting the value $L_1/L$ in equation (2), contact stiffness is determined. The main error in estimation of $k^*$ is due to uncertainty in geometry of tip.

Finally, in order to determine indentation modulus, the contact between tip and sample surface is considered by Hertzian contact mechanism [47]. With Hertzian contact consideration for spherical tip & flat sample surface and for a given static load for dominating elastic restoring forces, normal contact stiffness is given by

$$k^* = \sqrt[3]{6E^{*2}RF_N} \tag{7}$$

Where $E^*$ is reduced Young's modulus of elasticity and $E^*$ is related to the indentation modulus of tip and sample by following relation

$$\frac{1}{E^*} = \frac{1}{M_t} + \frac{1}{M_s} \tag{8}$$

Here $M_t$ & $M_s$ are indentation modulus of tip and sample respectively. Hertzian contact radius (nm) is given as following

$$a = \sqrt[3]{3F_N R / 4E^*} \tag{9}$$

From equations (7) and (9), following can be deduced

$$a = \frac{k^*}{2E^*} \quad (10)$$

To make elastic forces dominant, stiff cantilever is used and high static load $F_N$ is applied. For indentation modulus measurements using AFAM, a reference material with known indentation modulus, usually an amorphous or a single-crystal material with known orientation, is used to derive the indentation modulus of the test sample using a relation which follows Eq. (7) [46]

$$\frac{E_s^*}{E_r^*} = \left(\frac{k_s^*}{k_r^*}\right)^n \quad (11)$$

Where $k_s, k_s, k_r$ & $E_s^*, E_r^*$ are normal contact stiffness and reduced modulus of elasticity for test and reference samples respectively. Here n=3/2 for spherical tip in contact with flat surface and n=1 for flat tip in contact with flat sample. By combining equation (8) and (11), indentation modulus of sample can be expressed as following.

$$\frac{1}{M_s} = \frac{1}{M_r}\left(\frac{k_r^*}{k_s^*}\right)^n + \frac{1}{M_t}\left[\left(\frac{k_r^*}{k_s^*}\right)^n - 1\right] \quad (12)$$

Here indentation modulus of probe tip is assumed as of bulk of tip material. Even manufacturers do not measure indentation modulus of tip and only quote reported values. In our case it is even more ambiguous because it is DLC coated tip. Such assumption leads to the major uncertainties in the determination of indentation modulus of test sample and uncertainty can be ≥ 20%. Uncertainty in indentation modulus of tip could be reduced by using dual reference method and indentation modulus of sample using two references can be written as following [49]

$$M_s = \frac{\left(k^*_{r1}/k^*_{r2}\right)^n - 1}{\left(k^*_{r1}/k^*_s\right)^n \left[\frac{1}{M_{r2}} - \frac{1}{M_{r1}}\right] + \left(k^*_{r1}/k^*_{r2}\right)^n \left[\frac{1}{M_{r1}} - \frac{1}{M_{r2}}\right]} \quad (13)$$

Here $k^*_{r1}, M_{r1}$ and $k^*_{r2}, M_{r2}$ are contact stiffness & indentation modulus of reference (1) and reference (2) respectively. Dual reference method is insensitive to tip geometry or is marginally affected by tip geometry [49]. Dual reference method permits the calculations of the indentation modulus of the tip and is given by following relation [49].

$$M_T = \frac{\{(k^*_{r1})^n - (k^*_{r2})^n\} M_{r1} M_{r2}}{(k^*_{r2})^n M_{r1} - (k^*_{r1})^n M_{r2}} \quad (14)$$

In this way, by choosing two suitable references of known indentation modulus, the indentation modulus of tip can be determined with an uncertainty as low as 1%.

## 4. Results and Discussion

Fundamental and first free resonance frequencies for used DLC coated cantilever probe are shown in figure 2(a). For computing contact stiffness of test thin films, accurate value of stiffness or spring constant ($k_c$) of cantilever is required. Cantilever probe manufacturers do not quote exact value of $k_c$ for individual probe but quote a range of possible values. In order to determine $k_c$ accurately of used probe, fundamental free resonance spectra has been fitted using formulation proposed by Sadar and methodology is known as Sadar normal method. Fundamental free resonance frequency for used cantilever probe is 193.8 kHz and computed $k_c$ is 8.7 N/m. $1^{st}$ and $2^{nd}$ contact resonance frequency spectra have been measured for entire thin films. To keep parity in normal contact resonance measurements, same normal static load of

1919 nN has been applied on cantilever tip. As mentioned in section.3, if adhesion forces between cantilever tip and test sample, while being in contact, are comparable to normal static load, Hertzian contact mechanism fails. In this context, normal static load has been estimated from force-distance spectroscopy in AFM measurements for entire thin films. Force distance curves for sample SAMP-4 are shown in Fig. 4. Adhesion forces have been determined from pull-off region of force distance curves for entire samples. Magnitude of such forces is 24.5 nN for SAMP-4, and lies between 15 to 30 nN for entire films. To eliminate the effect of such non-linear adhesive forces, a sufficiently high normal static load of 1919 nN has been applied on cantilever tip. In fig. 2(b), $1^{st}$ & $2^{nd}$ contact resonances have been plotted as a function of applied static load. Resonance curves depict a Lorentzian shape and both amplitude & frequency of contact resonances increase with static load. Such behavior of contact resonances with static load follows flexural vibration theory [30]. Contact stiffness determined using characteristic equation (2) for $1^{st}$ & $2^{nd}$ contact resonances have been computed numerically for $L_1/L$ ranging from 0.85 to 0.98. The actual tip position and contact stiffness have been found by using equation (6), in which contact stiffness obtained from $1^{st}$ and $2^{nd}$ contact resonance frequencies are equated. Computed contact stiffness values for Si (100), BK7 glass, Sapphire and obliquely deposited $HfO_2$ thin films are listed in table 1 & 2 with corresponding $L_1/L$ values. Actual tip position lies between ($L_1/L$) 0.94 to 0.95 and normal contact stiffness lies between 638 to 1648 N/m for such films. Ratio $k^*/k_c$ varies from 73 to 189, which is close to 100 and the ratio in this range is considered as low. If ratio becomes of order of 1000 or exceeds 1000, it is considered as high. Consequently; $k^*/k_c$ ratio being low as described in reference [36], the lateral forces between cantilever tip and test sample are negligible and the influence of lateral contact stiffness on contact resonance frequencies is not remarkable and can be neglected. Contact stiffness plots for

1st and 2nd contact resonance frequencies as a function of effective tip position ($L_1/L$) for $HfO_2$ thin film SAMP-1, SAMPL-5, BK7 glass and Si(100) are shown in fig.3. Intersection of two curves gives contact stiffness and actual tip position. Uncertainty in normal contact stiffness calculations is around 0.5%; mainly due to uncertainly in contact resonance frequencies. It can also be noted that contact radius for a load of 1919 nN, as calculated from equation (10), is 5-8 nm for entire thin films and thickness of films as listed in table.3 is between 361 to 629 nm. It is worth to note that film thickness values are significantly higher than *3a*, which is generally accepted as the minimum thickness for neglecting stresses produced by film-substrate interface [28] and hence effect of substrate elastic properties on contact stiffness measurement of thin film samples can be neglected. Since the value of Poisson's ratio is not known for thin film, we have evaluated indentation modulus rather than Young's modulus of elasticity. Dual reference method as discussed in introduction has been adopted to determine indentation moduli of thin films. Although, in dual reference method, determination of indentation modulus is insensitive or very less sensitive to the geometry of cantilever tip, we have considered both spherical and flat tip and have taken mean of two cases. In such consideration, uncertainty is further reduced. As described in reference [49], uncertainty in dual reference method to determine indentation modulus is least when contact stiffness of two references bracket the stiffness of test sample. Consequently, for SAMP-1 & SAMP-2, BK7 glass & Si (100) references whereas for SAMP-3, SAMP-4 and SAMP-5; Si (100) & Sapphire references have been chosen. Equation (13) has been adopted to compute indentation moduli and values are listed in table.2. Indentation moduli for BK7 glass, Si (100) and sapphire has been taken from references. From table.2, it is clear that indentation modulus is least for GLAD $HfO_2$ with a mean value of 42 GPa and increases with deposition angle with highest value 221 GPa for SAMP-5, deposited at normal angle. K. Tapily et al., [50] have also reported a similar modulus of elasticity of 220 ± 40 GPa, measured by nano-indentation for $HfO_2$ thin film grown by atomic layer deposition. Sources of uncertainly in the determination of indentation moduli are uncertainty in tip geometry and indentation moduli of references. In present case, uncertainty in indentation moduli is less than 6% except SAMP-1. For SAMP-1, uncertainty is 16.7% and such a high uncertainty is the consequence of lower normal contact stiffness value of SAMP-1 than both the references BK7 glass and Si (100). Normalized grazing angle X-ray reflectivity spectra for entire $HfO_2$ thin films are shown in fig.5. To determine film density and RMS surface roughness, experimental spectra have been simulated with theoretical formulation by $\chi^2$ minimization using open source code IMD under XOP software package [51]. Density and RMS surface roughness values estimated through GIXR data analysis are listed in table.3. It is obvious from table.3 that RMS roughness is highest for SAMP-1 with a value 21 Å and is between 5-7 Å for rest of the thin films. High roughness for SAMP-1 is the result of highest shadowing effect among entire film during growth. Cross-sectional morphology of such films deposited has been recorded by FESEM and is presented in Fig.6. For SAMP-1, tilted columns are fully matured and measured tilt angle from cross-sectional morphology is 55 degree. It can be seen from FESEM images that void fraction is highest in SAMP-1 and it decreases as the angle α decreases. Column tilt angle (β) and film thickness measured from cross-sectional SEM morphology for entire thin films are listed in table.3. We have also calculated column tilt angle (β) and film density (ρ) theoretically using the ballistic growth model with limited ad-atom diffusion, proposed by Tait et al., [5]. As per model,

$\beta = \alpha - \arcsin\left(\frac{1-\cos\alpha}{2}\right)$ and $\rho = \rho_0\left(\frac{2\cos\alpha}{1+\cos\alpha}\right)$; where $\rho_0$ is the density of film deposited at α

=0. These theoretical formulations only give values to first approximation and actual values may differ significantly from theoretical. At high oblique angle, measured β matches with ballistic model whereas at smaller deposition angle, ballistic model fails to match with measured column tilt angle. It is well reported that ballistic model best explains such nanostructure thin films at large deposition angles [5, 6]. Moreover, in real films, chamber pressure, rate of deposition and non-monotonic behavior also perturb the growth and tilt angle of columns. Indentation modulus and density of films are plotted in Fig.7 as a function of angle of deposition and follow a parabolic decreasing trend. Cross-sectional FESEM morphology also depicts an increase in film porosity with deposition angle. Decrease in film density with α is the consequence of increase in film porosity due to tilted columnar microstructure (shadowing effect). Increase in porosity of the films with α leads to the decrease in indentation modulus or elastic modulus of the films. This type of trend is also reported and explained earlier [52-54].

## 5. Conclusion

A set of $HfO_2$ thin films have been deposited on Si (100) at angles 80, 68, 57, 40 and 0 degree by reactive electron beam evaporation. Such obliquely deposited films are known to be porous and their mechanical and environmental stability is an issue. In order to assess mechanical stability of such thin films, indentation modulus of entire thin films has been estimated by AFAM technique. Films have also been tested for their micro-structural properties by FESEM and GIXR measurements. FESEM cross-sectional morphology depicts very high porosity in thin film deposited at glancing angle (80 degree) & tilted columnar growth and porosity decreases as the angle of deposition decreases. Similar variation of film density has been achieved through GIXR measurements. It is observed that indentation modulus is least with a value of 42 GPa for GLAD film and highest for normally deposited film with value 221 GPa. It is concluded that decreasing trend of indentation modulus of $HfO_2$ thin films is an attribute of increase in porosity with angle of deposition and porosity variation is the consequence of change in microstructure of films due change in growth conditions.

**Caption of Figures:**

Fig.1 (a): Schematic of AFAM set-up.

Fig.1 (b): Simplified cantilever tip-sample contact interaction

Fig. 2(a): fundamental and $1^{st}$ overtone free resonance frequencies of DLC coated cantilever probe.

Fig. 2(b): Plot of $1^{st}$ and $2^{nd}$ tip-sample contact resonance frequencies as a function applied static load for thin film SAMP-5.

Fig. 3: Plot of contact stiffness for $1^{st}$ and $2^{nd}$ contact resonance frequencies as a function of $L_1/L$ for SAMP-2, SAMP-5 & glass and Si (100) references.

Fig. 4: Plot of force-distance curves taken by AFM for SAMP-4. In inset pull-off curve is highlighted to determine adhesion force.

Fig. 5: Grazing incidence X-ray reflectivity curves of thin films SAMP-1, SAMP-2, SAMP-3, SAMP-4 and SAMP-5.experimental and theoretically simulated curves for SAMP-1 are also shown.

Fig. 6: FESEM cross-sectional morphology for thin films SAMP-1, SAMP-2, SAMP-3, SAMP-4 and SAMP-5.

Fig. 7: variation indentation modulus and density obliquely deposited HfO2 thin films at different angles.

**Fig. 1 (a):**

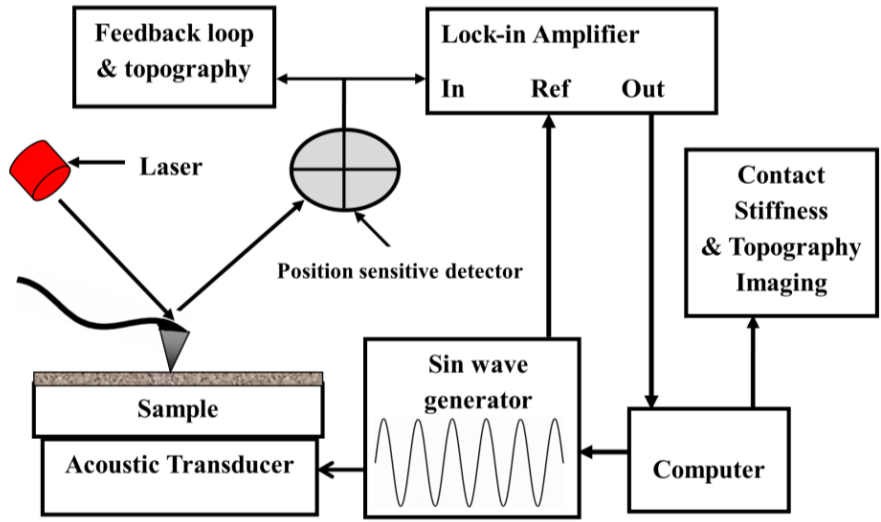

**Fig. 1 (b):**

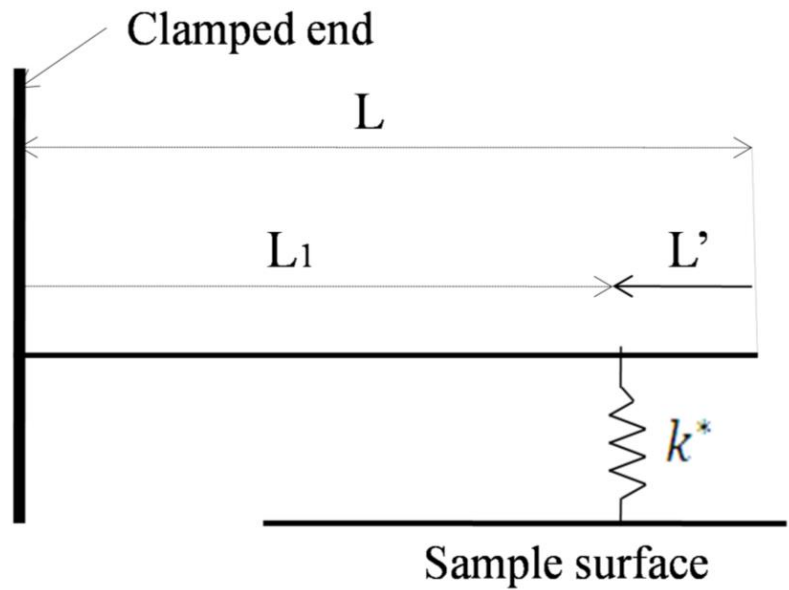

**Fig. 2(a):**

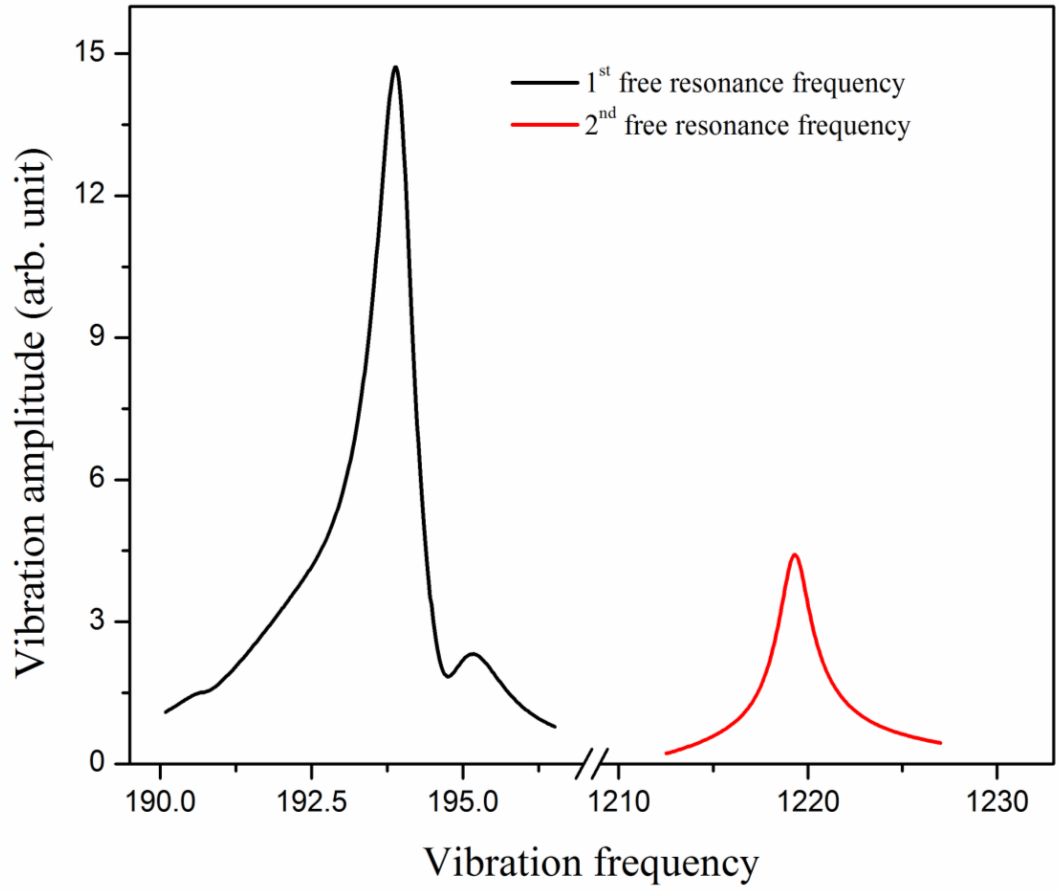

**Fig. 2(b):**

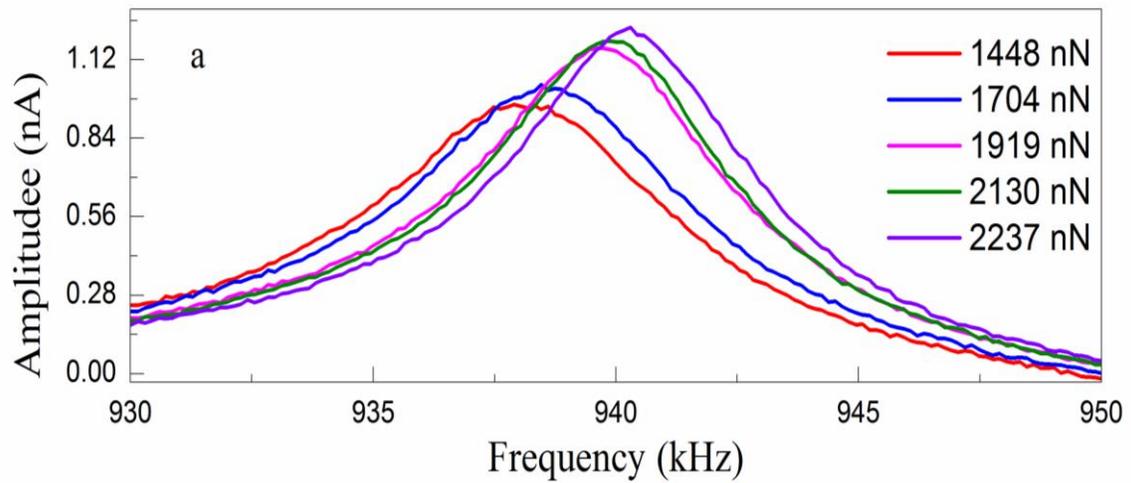

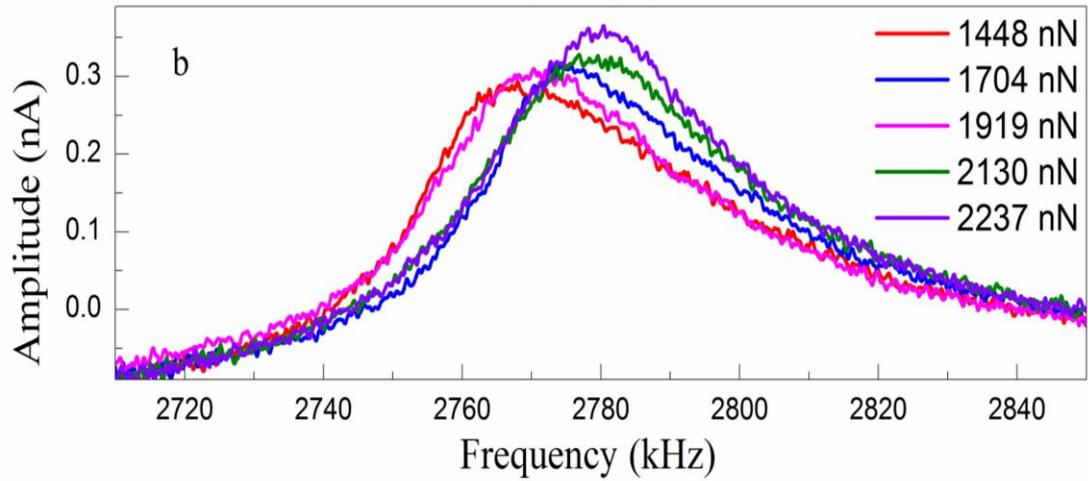

**Fig. 3:**

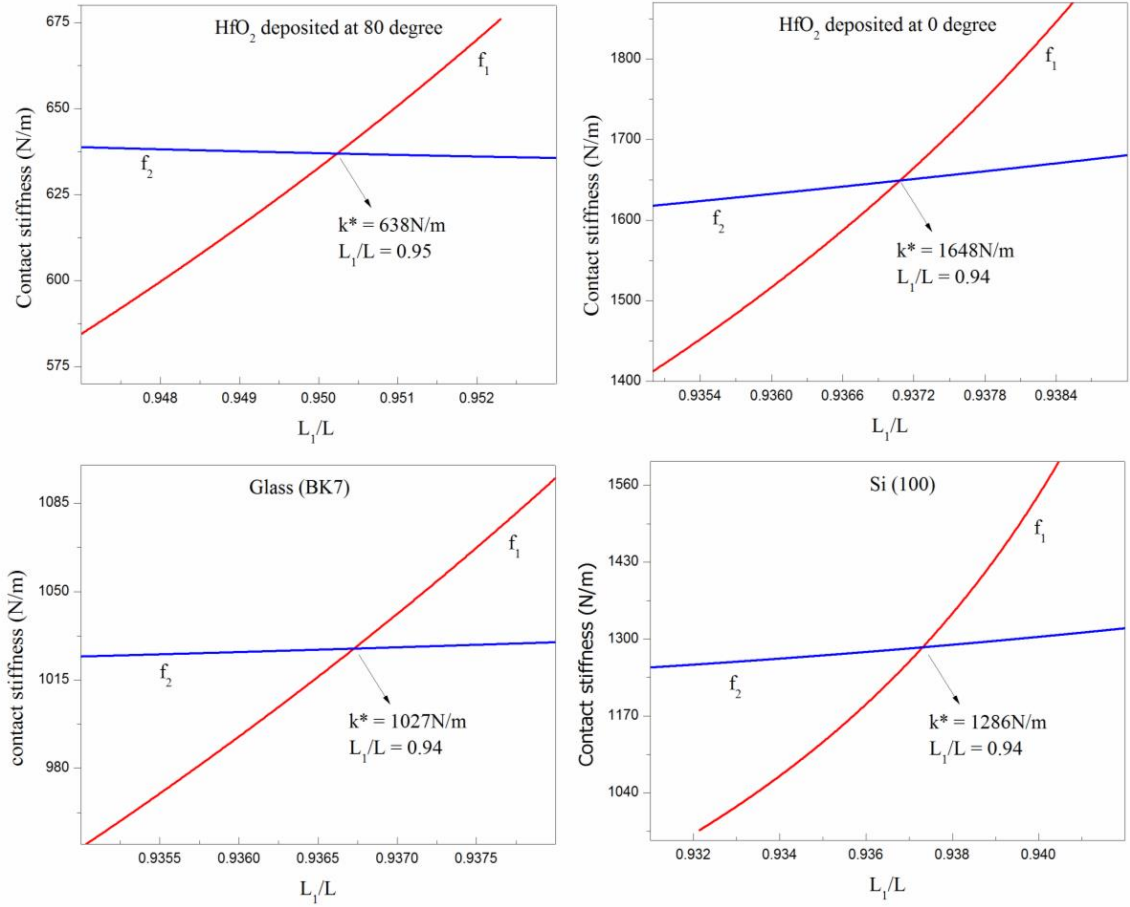

**Fig. 4:**

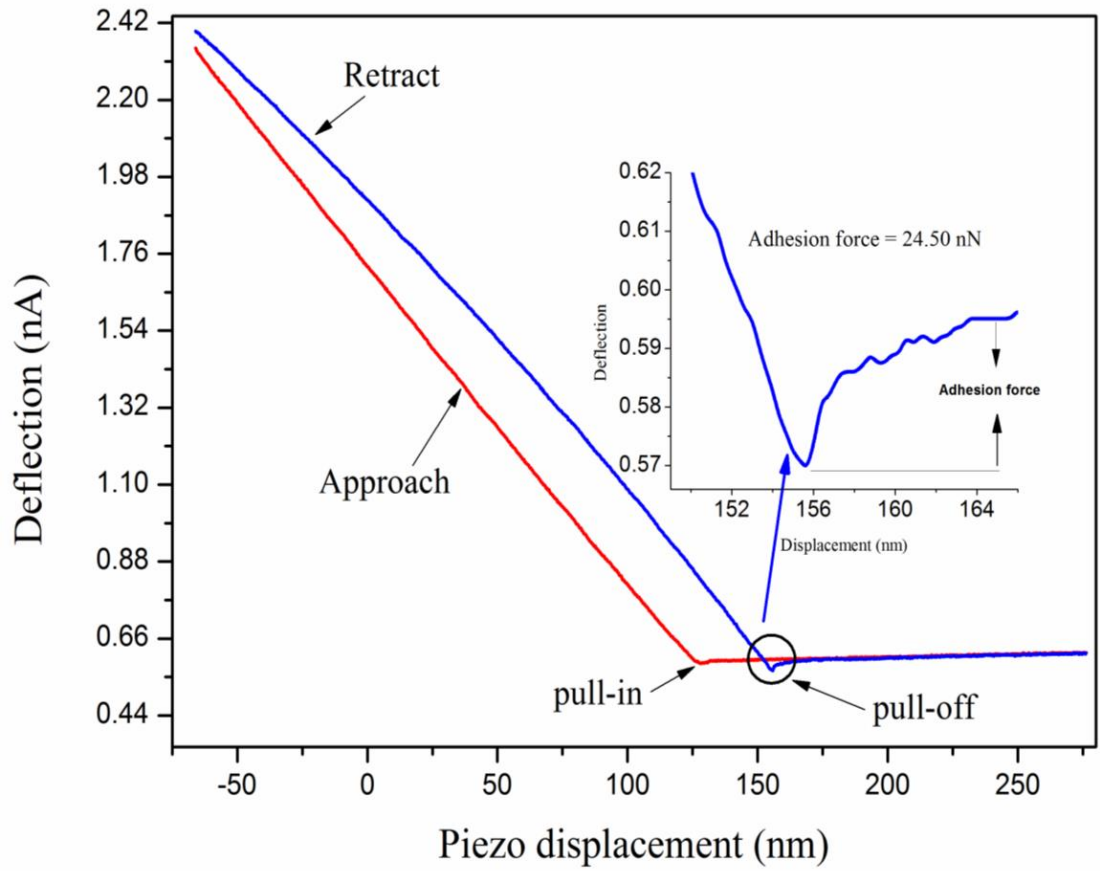

Fig. 5:

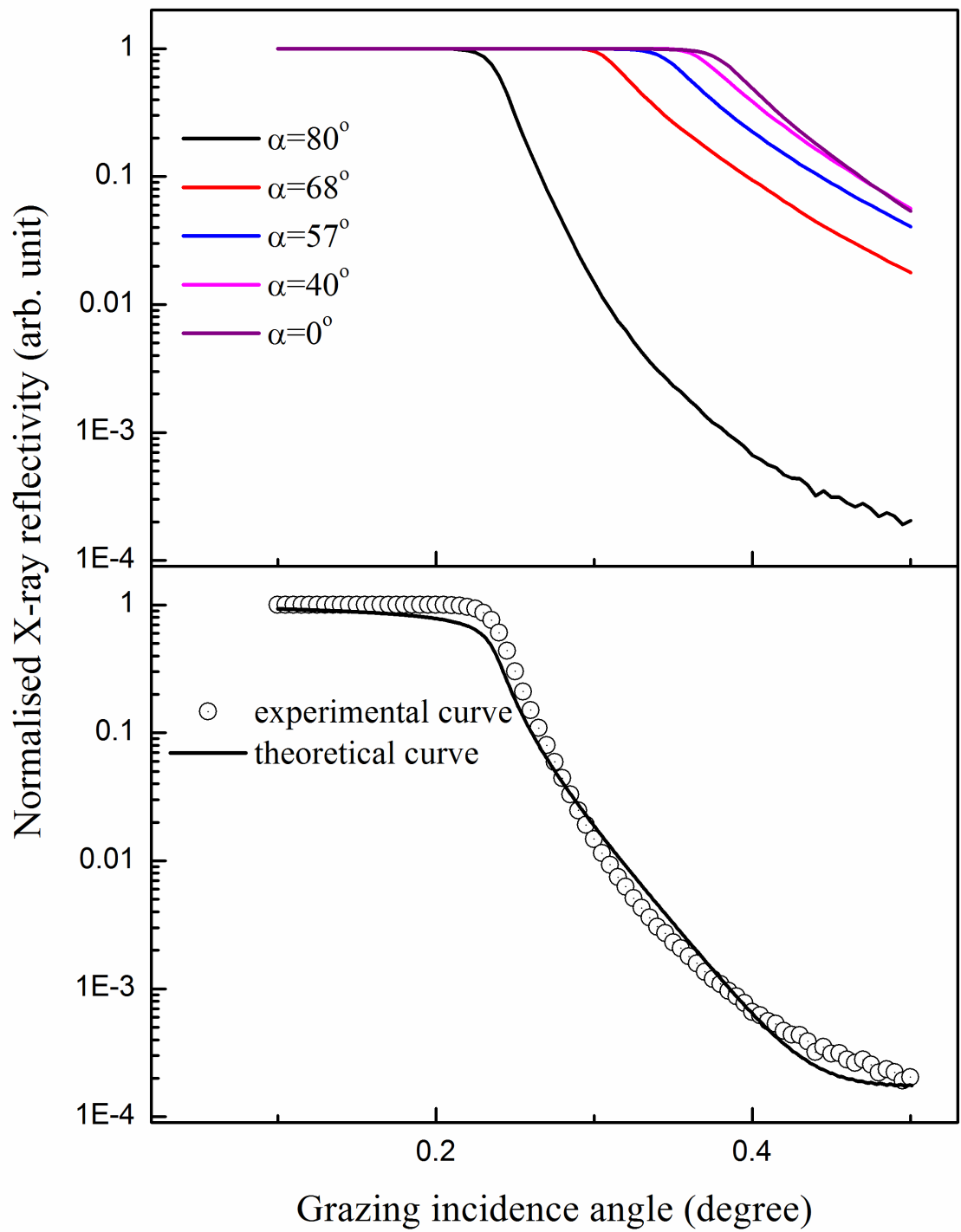

**Fig. 6:**

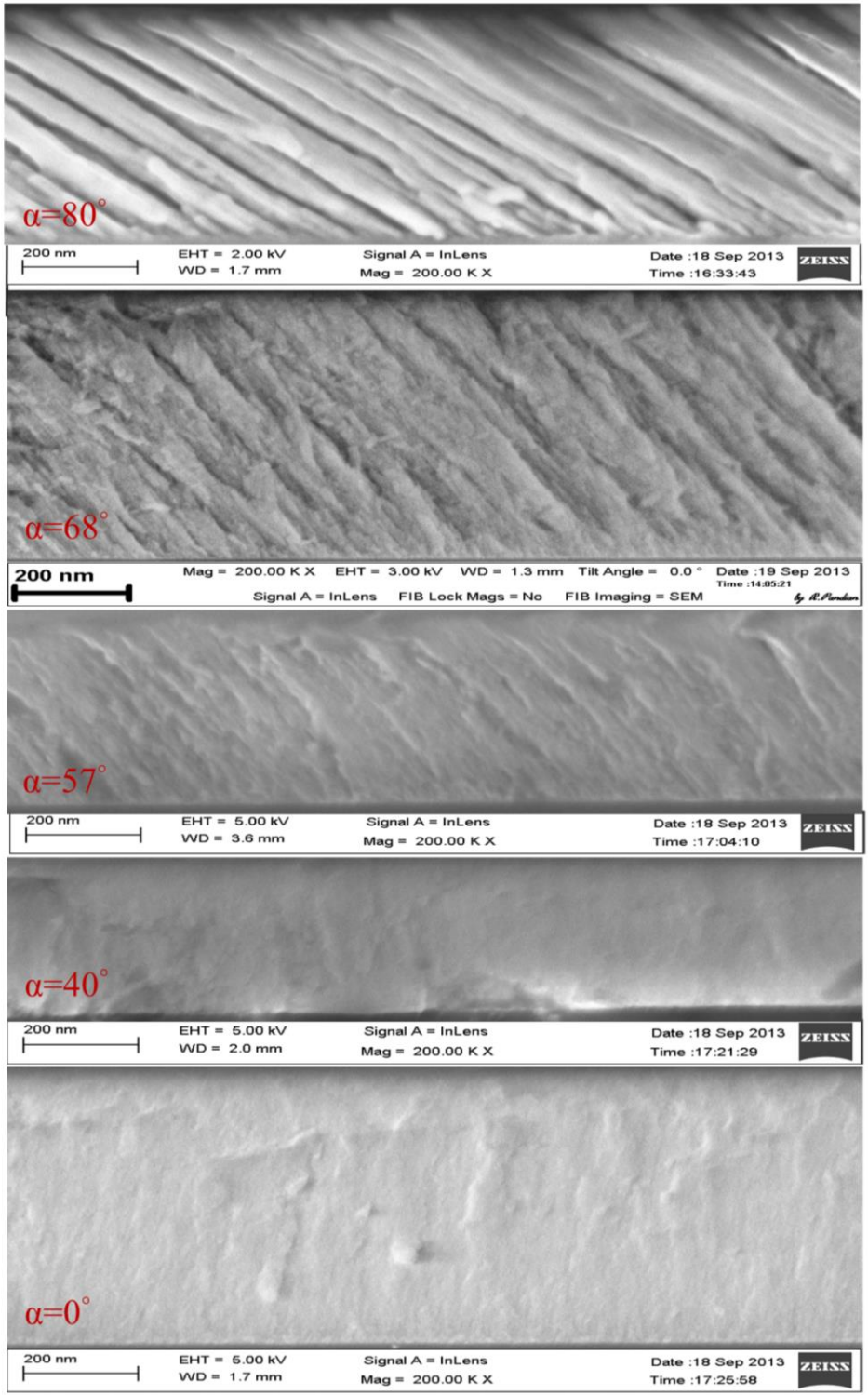

**Fig. 7:**

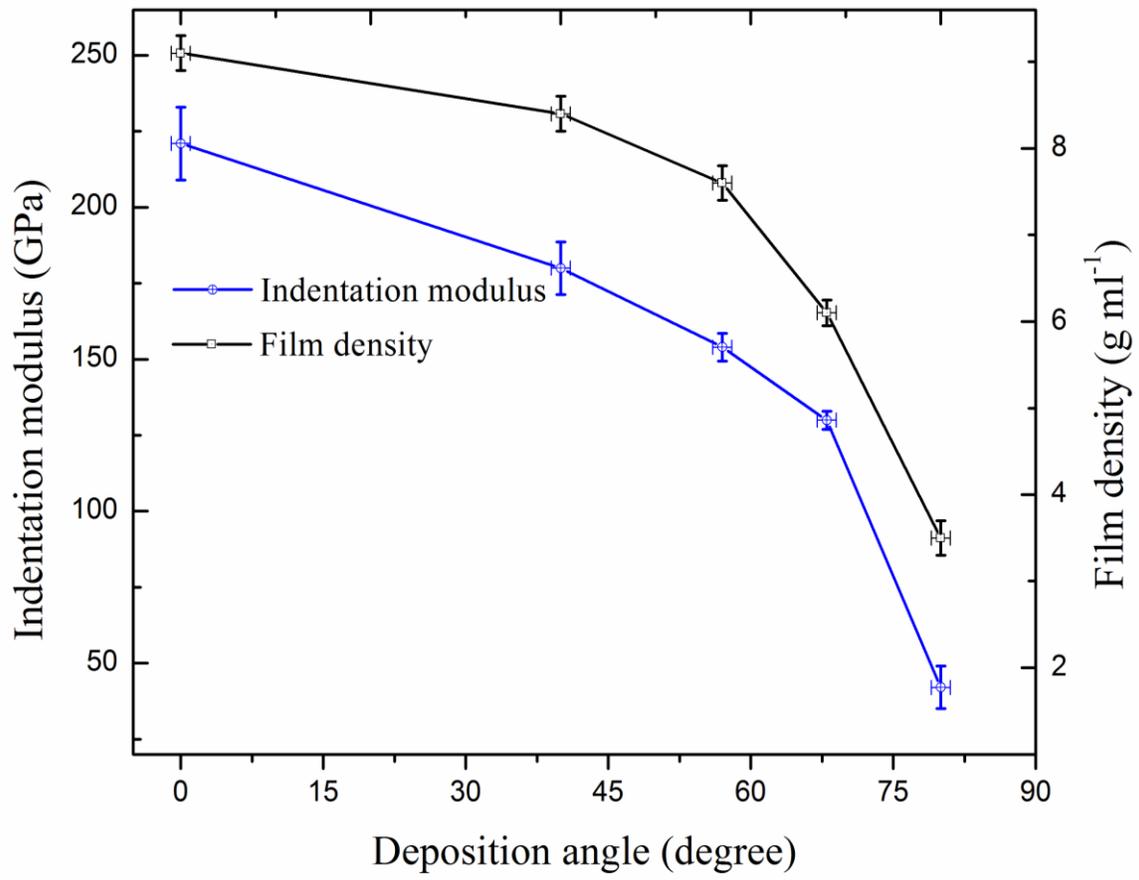

| 42Name of sample | 1$^{st}$ contact resonance frequency (kHz) | 2$^{nd}$ contact resonance frequency (kHz) | Contact stiffness (N/m) | Ratio ($L_1/L$) | Indentation modulus (GPa) |
|---|---|---|---|---|---|
| Si (100) | 935 | 2624 | 1286 | 0.94 | 135 ± 3 |
| BK7 Glass | 928 | 2473 | 1025 | 0.94 | 89 ± 2 |
| Sapphire | 945 | 2860 | 2113 | 0.94 | 420 ± 20 |

**Table.1:** Contact resonance frequencies, contact stiffness and indentation modulus of Si (100), BK7 and Sapphire references.

| Sample name | 1$^{st}$ Contact Resonance Frequency (kHz) | 2$^{nd}$ Contact Resonance Frequency (kHz) | Contact Stiffness (N/m) | Ratio ($L_1/L$) | Indentation Modulus (GPa) | Uncertainty (%) |
|---|---|---|---|---|---|---|
| SAMP-1 | 877 | 2131 | 638 | 0.95 | 42 ± 7 | 16.7 |
| SAMP-2 | 909 | 2573 | 1260 | 0.95 | 130 ± 3 | 2.3 |
| SAMP-3 | 940 | 2670 | 1379 | 0.94 | 154 ± 4.6 | 3.0 |
| SAMP-4 | 941 | 2715 | 1496 | 0.94 | 180 ± 8.7 | 4.8 |
| SAMP-5 | 941 | 2760 | 1648 | 0.94 | 221 ± 12 | 5.4 |

**Table.2:** Contact resonance frequencies, contact stiffness and indentation modulus of HfO$_2$ thin films deposited at different angles.

| Sample name | Film deposition angle (α) Degree | Measured Column tilt angle (β) in Degree | Theoretical Column tilt angle (β) in Degree | Measured Film density (g/cc) | Theoretical film density (g/cc) | Film thickness measured by FESEM | Film RMS roughness measured by GIXR (Å) |
|---|---|---|---|---|---|---|---|
| SAMP-1 | 80 ± 1 | 55 ± 2 | 55.1 ± .5 | 3.5 ± 0.2 | 2.9 ± 0.3 | 531 ± 2 | 21 |
| SAMP-2 | 68 ± 1 | 37 ± 2 | 49.8 ± 5 | 6.1 ± 0.2 | 5.0 ± 0.2 | 590 ± 2 | 5 |
| SAMP-3 | 57 ± 1 | 33 ± 2 | 43.8 ± .6 | 7.6 ± 0.2 | 6.4 ± 0.2 | 488 ± 2 | 6 |
| SAMP-4 | 40 ± 1 | 12 ± 2 | 33.2 ± .7 | 8.4 ± 0.2 | 7.9 ± 0.1 | 361 ± 2 | 7 |
| SAMP-5 | 0 ± 1 | 0 | 0 ± 1 | 9.1 ± 0.2 | 9.1 | 629 ± 2 | 7 |

**Table.3**: results of FESEM, AFM and GIXR measurements